\documentclass[sigconf]{acmart}
\setcopyright{none}              \renewcommand\footnotetextcopyrightpermission[1]{}  
\AtBeginDocument{%
  }
\usepackage{graphicx}
\usepackage{algorithm,algorithmic}
\usepackage{enumitem}
\usepackage{pifont}
\usepackage{makecell}
\definecolor{salmon}{RGB}{250,128,114}
\definecolor{ForestGreen}{RGB}{34,139,34}

\copyrightyear{2026}
\acmYear{2026}
\setcopyright{cc}
\setcctype{by-nc-nd}
\acmConference[GLSVLSI '26]{Great Lakes Symposium on VLSI 2026}{June 22--24, 2026}{Canandaigua, NY, USA}
\acmBooktitle{Great Lakes Symposium on VLSI 2026 (GLSVLSI '26), June 22--24, 2026, Canandaigua, NY, USA}
\acmDOI{10.1145/3787109.3815309}
\acmISBN{979-8-4007-2431-2/2026/06}

\begin{document}

\newcommand{\M}{µ-ORCA}
\title{\M: Optimizing Acceleration for Microsecond-Scale\\ Deep Neural Network Inference on ACAP
}

\author{Shixin Ji}
\affiliation{%
  \institution{Brown University 
  \city{Providence}
  \country{USA}
  }}
\email{shixin_ji@brown.edu}

\author{Jinming Zhuang}
\affiliation{%
  \institution{Brown University 
  \city{Providence}
  \country{USA}
  }}
\email{jinming_zhuang@brown.edu}

\author{Zhuoping Yang}
\affiliation{%
  \institution{Brown University 
  \city{Providence}
  \country{USA}
  }}
\email{zhuoping_yang@brown.edu}

\author{Xingzhen Chen}
\affiliation{%
  \institution{Brown University 
  \city{Providence}
  \country{USA}
  }}
\email{xingzhen_chen@brown.edu}

\author{Wei Zhang}
\affiliation{%
  \institution{Brown University 
  \city{Providence}
  \country{USA}
  }}
\email{wei_zhang6@brown.edu}

\author{Peipei Zhou}
\affiliation{%
  \institution{Brown University 
  \city{Providence}
  \country{USA}
  }}
\email{peipei_zhou@brown.edu}


\begin{abstract}
Heterogeneous reconfigurable platforms with tensor cores, such as AMD ACAP, are increasingly adopted for deep neural network (DNN) inference due to their high throughput and flexibility.
However, their suitability for microsecond-scale inference on small problem sizes remains underexplored.
In jet-tagging applications in high-energy physics, inefficient on-chip communication and large inter-layer latency prevent existing frameworks from meeting the 1-$\mu$s latency budget.
Moreover, hardware overheads such as synchronization and VLIW processor prologue are often overlooked, making it infeasible to optimize accelerators correctly.
To address these problems, we propose~\M, a customized heterogeneous accelerator framework for ultra-low-latency model inference.
\M~enables direct inter-layer communication between DNN layers on the AIE array, instead of using shared memory tiles or FPGA fabric.
Moreover, a 512-bit/cycle cascade connection is applied instead of a 32-bit/cycle DMA connection.
\M~also provides an overhead-aware performance model that adapts to different NN layer sizes, and conducts design space exploration to optimize end-to-end latency.
\M~supports MLP and DeepSets models with non-MM kernels, including bias, ReLU, and global aggregation on AIE.
We evaluate \M~on the AMD ACAP VEK280 platform.
Experimental results show that~\M~achieves average latency reduction of $>$1.70$\times$ and $>$1.83$\times$ compared with different state-of-the-art ACAP frameworks, and achieves 0.93 $\mu$s latency for a 6-layer real-world DeepSets model, satisfying the latency budget.
We open source~\M~at \url{https://github.com/arc-research-lab/u-ORCA}.
\end{abstract}



\pagestyle{plain}
\maketitle

\section{Introduction}
\label{sec: introduction}

\begingroup
\renewcommand\thefootnote{}
\footnotetext{This paper has been accepted at GLSVLSI 2026}
\endgroup

Deep neural networks (DNNs) are widely used in many application areas, and ultra-low-latency inference is becoming increasingly important.
A representative example is jet tagging in particle physics, where DNNs assist in filtering the massive data streams generated by particle colliders. The collider operates at 40 MHz~\cite{cms-l1-trigger} and produces hundreds of terabytes of data per second. The trigger system must decide which data to retain within less than $10~\mu s$~\cite{Next-gen_trigger}, leaving only a small time budget for neural network inference, e.g., $3.2~\mu s$~\cite{Taking-a-closer-look-at-LHC}.
General-purpose AI accelerators such as GPGPUs are not well-suited for ultra-low-latency inference due to their architecture. For example, GPUs load inputs and store outputs for each layer through main memory, and a single memory load operation can take around 200 ns~\cite{Demystifying_the_Nvidia_Ampere}. Such overhead accumulates across layers and becomes significant under $\mu$s-level latency constraints.
In contrast, FPGA accelerators have shown effectiveness for ultra-low-latency inference. 
They usually use lookup tables as multipliers, fully unroll computations to exploit massive parallelism, and keep all intermediate data on-chip to eliminate off-chip memory latency.
However, FPGA has limited on-chip resources, which restricts the model size, and usually requires high sparsity and low-bitwidth quantization to the model.
As a result, model performance, such as accuracy, may need to be compromised.

Heterogeneous accelerators leveraging vector units, such as AMD ACAP accelerators, are promising for ultra-low-latency NN inference due to their high computation capability and flexibility.
However, existing ACAP accelerator frameworks primarily target large-scale models and throughput optimization, limiting their support for microsecond-scale inference latency.
First, inter-layer on-chip communication becomes the bottleneck.
Most existing frameworks~\cite{zhuang2023charm,zhuang2023automm,charm2024trets,ssr,eq_vit_tcad,fpga25aries} follow a layer-wise execution model, using the AIE array for computation and the FPGA fabric for on-chip data reuse.
In this model, each layer's input must be transferred from the FPGA fabric to the AIE array, and its output must be written back after computation.
These transfers rely on DMA communication with only 32-bit/cycle transfer bandwidth.
Consequently, for tiny ultra-low-latency models, inter-layer communication dominates end-to-end latency, preventing existing architectures from operating efficiently at the microsecond scale.
In addition, existing design space exploration methods are insufficient as hardware overheads are not precisely characterized.
To achieve 1~$\mu$s overall latency, each layer has only about 100~ns of latency budget.
However, control logic, synchronization, and VLIW processor prologue before computation can introduce tens to hundreds of nanoseconds of overhead, significantly affecting end-to-end latency.
Without accurately quantifying these hardware costs, the design space for microsecond-scale inference cannot be effectively explored.

To address these problems, we propose~\M~, a customized heterogeneous accelerator framework for ultra-low-latency model inference.
~\M~ implements all DNN layers within the AIE array and enables direct communication between consecutive layers without involving the FPGA fabric or shared memory tiles.
Furthermore, ~\M~ leverages the 512-bit cascade connection for inter-layer data transfer, significantly reducing the on-chip communication latency between layers.
~\M~also quantifies hardware operations and proposes a fine-grained, overhead-aware performance model.
A design space exploration (DSE) is conducted to optimize the latency automatically.
~\M~supports both MLP and DeepSets models, which are widely used in jet-tagging applications. 
In addition to matrix-multiplication(MM) layers, high-performance implementations of non-matrix-multiplication layers, including global aggregation in DeepSets, bias addition, and ReLU, are designed and implemented on AIEs.
We evaluate~\M~across various scales and workloads. Experimental results show that~\M~achieves more than 1.70$\times$ and 1.83$\times$ latency reduction for synthetic MLP and realistic workloads, respectively, and achieves lower than the 1-$\mu$s budget.
Also, the global aggregation layer reaches $>$2.8$\times$ reduction over the baselines.
Furthermore,~\M's performance model shows an average estimation error of 1.1\% on single AIE and 15.6\% on the end-to-end models, whereas baseline performance models exhibit $>$25.5\% on single AIE.

In summary, our contributions are as follows:
\vspace{-5pt}
\begin{itemize}[leftmargin=*]
\item \textbf{Advanced on-chip inter-layer communication:}
\M~enables NN layers implemented on AIE tiles to communicate directly and leverages the 512-bit cascade connections for high-bandwidth inter-layer data transfer, significantly reducing inter-layer communication latency.
\item \textbf{Overhead-aware performance model and DSE:}
We develop a fine-grained performance model that quantifies both computation time and hardware overheads. Based on this model, a DSE is developed to optimize end-to-end inference latency.
\item \textbf{High performance non-MM kernel design:}
We design and implement high-efficiency non-MM kernels, including the global aggregation layer, enabling end-to-end inference within the AIE array for MLP and DeepSets models.
\end{itemize}

\section{Related Works}
\begin{table}[t]
\centering
\footnotesize
\caption{Architecture features of \M~and comparisons with prior works of AIE-based frameworks.}
\vspace{-5pt}

\begin{tabular*}{\linewidth}{@{\extracolsep{\fill}}l|c c c c c c@{}}
\hline
\textbf{Framework} & \textbf{IC} & \textbf{CIC} & \textbf{OPM} & \textbf{GA} & \textbf{BRA} & \textbf{PO} \\ \hline
\makecell[l]{CHARM~\cite{zhuang2023charm}\\autoMM~\cite{zhuang2023automm}\\CHARM2.0~\cite{charm2024trets}} & DDR & \textcolor{red}{\ding{55}} & \textcolor{red}{\ding{55}} & \textcolor{red}{\ding{55}} & \textcolor{red}{\ding{55}} & \textcolor{ForestGreen}{\ding{51}} \\ \hline
\makecell[l]{MAXEVA~\cite{maxeva}\\AMA~\cite{deng2024ama}\\GAMA~\cite{mhatre2025gama}} & - & \textcolor{red}{\ding{55}} & \textcolor{red}{\ding{55}} & \textcolor{red}{\ding{55}} & \textcolor{red}{\ding{55}} & \textcolor{red}{\ding{55}} \\ \hline
\makecell[l]{SSR~\cite{ssr}, \\EQ-ViT~\cite{eq_vit_tcad}} & PL & \textcolor{red}{\ding{55}} & \textcolor{red}{\ding{55}} & \textcolor{red}{\ding{55}} & \textcolor{red}{\ding{55}} & \textcolor{ForestGreen}{\ding{51}} \\ \hline
ARIES~\cite{fpga25aries} & PL & \textcolor{red}{\ding{55}} & \textcolor{red}{\ding{55}} & \textcolor{red}{\ding{55}} & \textcolor{red}{\ding{55}} & \textcolor{ForestGreen}{\ding{51}} \\ \hline
AIE4ML~\cite{danopoulos2025aie4ml} & \makecell[c]{shared\\mem tile} & \textcolor{red}{\ding{55}} & \textcolor{red}{\ding{55}} & \textcolor{red}{\ding{55}} & \textcolor{ForestGreen}{\ding{51}} & \textcolor{ForestGreen}{\ding{51}} \\ \hline
IRON~\cite{AMD_IRON} & - & \textcolor{red}{\ding{55}} & \textcolor{red}{\ding{55}} & \textcolor{red}{\ding{55}} & \textcolor{ForestGreen}{\ding{51}} & \textcolor{red}{\ding{55}} \\ \hline
\makecell[l]{\M\\(ours)} & direct & \textcolor{ForestGreen}{\ding{51}} & \textcolor{ForestGreen}{\ding{51}} & \textcolor{ForestGreen}{\ding{51}} & \textcolor{ForestGreen}{\ding{51}} & \textcolor{ForestGreen}{\ding{51}} \\ \hline
\end{tabular*}

{\raggedright
\scriptsize
\begin{tabular}{@{}l l@{}}
\textbf{IC}: Inter-layer Communication pattern. & \textbf{CIC}: Cascade-enabled Inter-layer Communication.\\
\textbf{OPM}: Overhead-aware Performance Model. & \textbf{GA}: Global Aggregation layer.\\
\textbf{BRA}: Bias and ReLU on AIE. & \textbf{PO}: Automatic Performance optimization.\\
\end{tabular}
}
\label{tab:feature_comparison}
\vspace{-20pt}
\end{table}
In this section, we summarize existing works in the ultra-low-latency DNN inference domain and compare them with~\M.

Various approaches have been proposed to optimize DNN inference on AMD ACAP architectures. A comparison of their key features is provided in Table~\ref{tab:feature_comparison}, with parameter abbreviations explained in the note below the table. In terms of communication, CHARM~\cite{zhuang2023charm}, autoMM~\cite{zhuang2023automm}, and CHARM~2.0~\cite{charm2024trets} load input data from and store output data to off-chip memory for each layer, while leveraging PL to improve data reuse and implement bias and non-linear operations without exploring on-chip forwarding. SSR~\cite{ssr} and EQ-ViT~\cite{eq_vit_tcad} transfer inter-layer intermediate data through on-chip communication and buffer all data and weights in the PL.
However, for each layer, the data is still transferred between the PL and the AIE array via PLIO.
MAXEVA~\cite{maxeva}, AMA~\cite{deng2024ama}, and GAMA~\cite{mhatre2025gama} aim to improve the computation efficiency for the matrix multiplication kernel without exploring inter-layer data movement.
AIE4ML~\cite{danopoulos2025aie4ml} leverages the global shared memory tiles of the AIE-ML architecture to store inter-layer intermediate data, while preloading weights into local memory tiles.
This design uses a 32-bit/cycle DMA connection for communication between AIE kernels, whereas \M~leverages the 512-bit/cycle cascade connection to transfer inter-layer intermediate data. By exploiting higher-bandwidth inter-AIE connections, \M~reduces communication latency, which is critical for meeting the 1~$\mu$s latency budget of medium-scale DNN models in scientific computing applications, such as jet tagging in particle physics experiments.
Other frameworks leveraging the Versal ACAP architecture~\cite{ma2026design,wang2025rsn, Nguyen2023spades, Bouaziz2025MonteCarlo,ji2026derca,yang2023aim} further demonstrate the effectiveness of ACAP across diverse workloads, while generally relying on DMA-based communication.
~\cite{fccm26_placement} discusses different placement methods for mapping DNN models onto the AIE array in AI PCs.
Several works have also focused on programming abstractions for AIE-based architectures, where communication is typically implemented using DMA and shared local memory.
ARIES~\cite{fpga25aries} proposes a Python-based programming model and automatically optimizes performance through MLIR-based compiler passes. However, it also relies on DMA for inter-layer data movement and does not fully exploit the high on-chip bandwidth available in the Versal ACAP architecture.
AMD IRON~\cite{AMD_IRON} also provides a Python-based frontend for programming Ryzen AI NPUs, offering a set of predefined kernels that users can compose to build complete applications. The high-bandwidth interconnects, such as the cascade connection, remain underexplored.
CHARM~\cite{zhuang2023charm}, SSR~\cite{ssr}, AMA~\cite{deng2024ama}, and~\cite{mhatre2025performance} build analytical models.
However, several hardware overheads, including VLIW processor prologue latency and DMA communication, are not modeled. For our target application with a stringent latency budget of 1~$\mu$s, these overheads cannot be ignored. As a result, these approaches are not well suited for ultra-low-latency scenarios, whereas~\M~maintains high accuracy across a broad range of model sizes.
Furthermore, directly supporting global aggregation layers on AIEs remains underexplored, whereas~\M~provides a high-performance design and implementation.

There are other works~\cite{LL-ViT, LUTNet, LogicNets, PolyLUT, schulte2025hls4ml, da4ml, cassidy2025reducedlut, LUTPIM} providing ultra-low-latency solutions for DNN models on FPGA platforms. These approaches primarily explore customized numerical data types and computation units implemented using lookup tables (LUTs) or DSPs on PL. Their efforts are orthogonal to~\M~, which focuses on ultra-low-latency inference for AIE architectures. 
Hardware-software co-design approaches, including DWN~\cite{DWN} and HGQ~\cite{HGQ}, have also been proposed. These methods are optimized for low-bitwidth quantized models, while~\M~targets dense INT8 models.
Several ultra-low-latency neural network designs have been implemented on FPGA fabrics, including DeepSets~\cite{odagiu2024ultrafast}, MLP-Mixer~\cite{sun2025fast}, and Transformer-based models~\cite{laatu2025sub}.
In~\M, we focus on MLP-based and DeepSets-based models. The exploration of hardware-software co-design techniques for ACAP is left for future work.

\vspace{-5pt}
\section{Motivation Examples and Design Challenges}
\label{sec: motivation examples}
In this section, we demonstrate the key design challenges and research opportunities in designing DNN accelerators with 1-$\mu$s scale latency. 
We first explain how~\M~reduces overall latency by improving on-chip data movement, and then describe the communication-computation trade-off enabled by the~\M~cascade communication pattern.

\vspace{-5pt}
\subsection{On-chip Data Communication}
\begin{figure}
    \centering
    \includegraphics[width=0.75\linewidth]{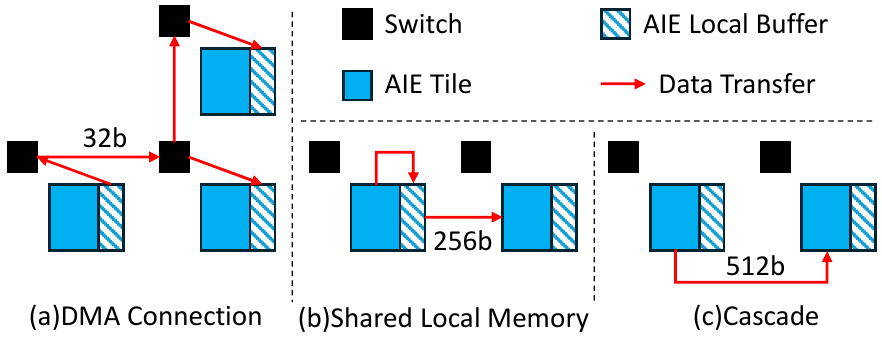}
    \vspace{-10pt}
    \caption{Data movement methods among AIE tiles.
    }
    \vspace{-10pt}
    \label{fig: communication_patterns}
\end{figure}
Figure~\ref{fig: communication_patterns} shows the three methods for data movement among different tiles within the AIE array.  
First, \textbf{DMA communication} (Figure~\ref{fig: communication_patterns}(a)) can be used, in which data is transferred via the network-on-chip (NoC) within the AIE array. 
An AIE tile can send data to an arbitrary tile using DMA, and multicasting is supported.
However, DMA has a constrained bandwidth of only 32 bits/cycle. 
In addition, adjacent AIEs can communicate via the \textbf{shared local memory} between them (\ref{fig: communication_patterns}(b)), offering 256 bits/cycle bandwidth.
Both DMA and shared local memory connection use the \texttt{IO\_buffer} API.
Lock-based Synchronization is needed for the buffers, and the destination AIE cannot access the data until the buffer is fulfilled, incurring hardware overheads in communication.
\textbf{Cascade communication} (\ref{fig: communication_patterns}(c)) can also be used, featuring 512 bits/cycle bandwidth.
In the AIE-ML architecture, one AIE can only cascade to its neighbor in the east or south.
The cascade is implemented as a FIFO in hardware, requiring almost no synchronization, and the overhead is trivial. 

\begin{figure}
    \centering
    \includegraphics[width=0.75\linewidth]{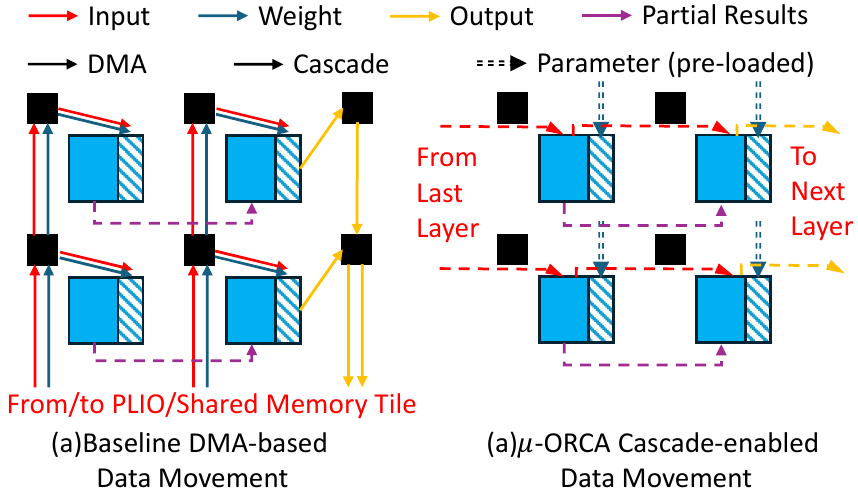}
    \vspace{-10pt}
    \caption{Data movement pattern of MM kernels.
    }
    \vspace{-15pt}
    \label{fig: motivation_example_data_movement}
\end{figure}

With a matrix-multiplication layer example, we show the optimization opportunities of on-chip data movement that existing works overlook.
As shown in Figure~\ref{fig: motivation_example_data_movement}, 4 different kinds of data will be transferred among the AIEs: input, weight, output of this layer, and the partial results that need to be accumulated along the reduced(K) dimension.
The baseline communication patterns shown in~\ref{fig: motivation_example_data_movement}(a) use DMA communication to load the input and weights from the FPGA fabric.
To enable parallelism along the reduced dimension, the partial results are transferred via cascade~\cite{zhuang2023charm,charm2024trets,zhuang2023automm,ssr,eq_vit_tcad,mhatre2025gama,danopoulos2025aie4ml} or shared local memory connection~\cite{maxeva}.
The kernel latency can be bounded by the low-bandwidth DMA communication.
Consider a $32\times32\times32$ INT8 MM kernel on 4 AIE-ML tiles with spatially unroll M and K dimensions by 2. 
In the ideal case, it takes only 32 cycles for computation with 256 MAC/cycle per AIE, but 64 and 128 cycles to load the input and weight, and 128 cycles to store the output.
As a result, the layer latency is no smaller than 288 cycles.

\M~solves this bottleneck via enabling direct communication between layers, and enabling the cascade for the intermediate data between layers.
As demonstrated in Figure~\ref{fig: motivation_example_data_movement}(b), the input is cascaded from the last layer, and the output is cascaded to the next layer directly, and the weight is preloaded to the local memory as runtime parameters.
It takes only 8 cycles for input or output, and the layer latency is only 48 cycles, reaching a 6x reduction ideally.

\vspace{-10pt}
\subsection{Communication-computation trade-off}
\begin{figure}
    \centering
    \includegraphics[width=0.75\linewidth]{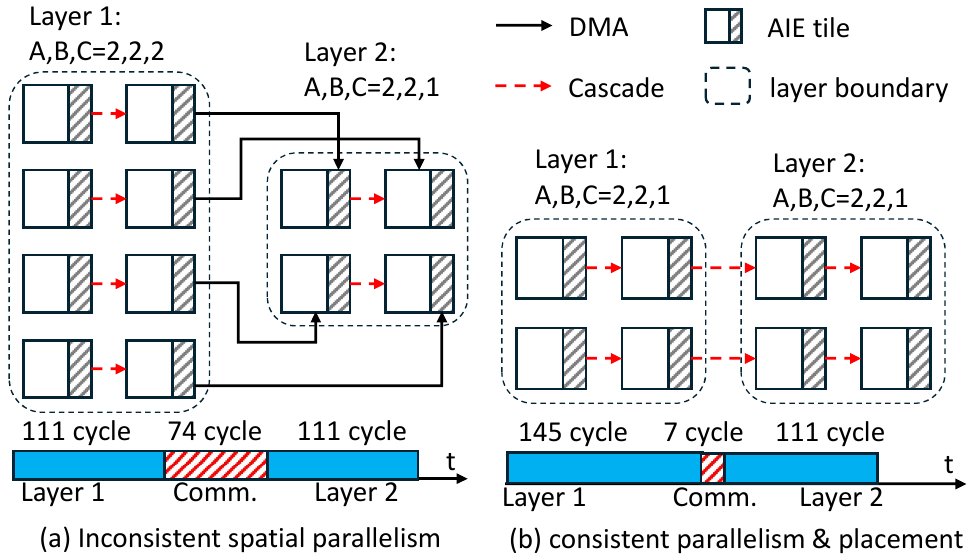}
    \vspace{-10pt}
    \caption{Two design points of consecutive MM layers with inconsistent and consistent workload partitions. For each layer, the workload is spatially partitioned by A, B, and C in the M, K, and N dimensions, respectively.
    }
    \vspace{-20pt}
    \label{fig: motivation_example_dse}
\end{figure}
The direct cascade connection enabled by~\M~poses new challenges to the design space exploration.
When deciding the spatial parallelism of consecutive layers, the parallelism not only affects the computation latency, but also the communication latency, as the communication amount and pattern change with the parallelism.
Besides, the high-bandwidth cascade communication can only be enabled within certain spatial partitions due to the topology constraint, which we discuss in detail in~\ref{sec: inter-layer communication}.
Here, we show the communication-computation trade-off via an illustrative example of two consecutive MM layers with shapes of $8\times64\times64$ and $8\times64\times32$, respectively.
Figure~\ref{fig: motivation_example_dse}(a) shows the design points optimizing for only the computation, where 8 and 4 AIEs are allocated for layers 1 and 2.
As the number of rows does not match, only DMA can be used, and the overall latency is 296 cycles.
Then, in Figure~\ref{fig: motivation_example_dse}(b), we reduce the parallelism of layer 1 to 4 AIEs, which increases layer 1 computation time to 145 cycles.
On the other hand, the row of layer 1 and layer 2 now matches, and the cascade connection is enabled, reducing the communication latency from 74 to 7 cycles.
As the communication latency is greatly reduced by enabling cascade, the overall latency is optimized from 296 cycles to 263 cycles.

\vspace{-10pt}
\section{\M~Architecture and Implementation}
In this section, we demonstrate the architecture and design parameters of~\M~at the single AIE layer level, then we explain the methodology of transferring intermediate data across layers.

\vspace{-10pt}
\subsection{Single AIE and Layer design}
We implement the AIE kernel function based on the AIE-ML API~\cite{aie-api} and intrinsics~\cite{aie-ml-intrinsic}.
One AIE kernel in~\M~supports loading one source operand via cascade or DMA, and the other via DMA or to be pre-loaded to AIE local memory as runtime parameters.
The partial or full results can be sent via cascade or DMA.

A single AIE kernel is performed in two phases of communication and computation.
When cascade is used, one layer has to send its output to non-neighboring AIE tiles by leveraging the tiles in between them; thus, each AIE will load from its upstream, store its own part of the data, and send it to the downstream kernel.
Detailed data movement pattern will be introduced in Section~\ref{sec: inter-layer communication}.
When DMA is used, the data can be multicasted and sent to the destination directly, and the communication phase is skipped.

During the computation phase, the output-stationary loop permutation is used.
The AIE vector processor performs matrix multiplication in a blocked manner, with each cycle processing a small matrix multiplication (MM) within a single instruction.
To enhance data reuse in the local memory, the \texttt{i} and \texttt{j} loops are unrolled by 2.
We denote the shape of one block as $B_M\times B_K \times B_N$, which is determined by the AIE architecture and the kernel bitwidth precision (e.g. $4\times8\times8$ in AIE-ML for INT8).
We also denote the total matrix multiplication shape for one AIE kernel as $H_1\times W_1 \times W_2$.

\begin{figure}
    \centering
    \includegraphics[width=0.85\linewidth]{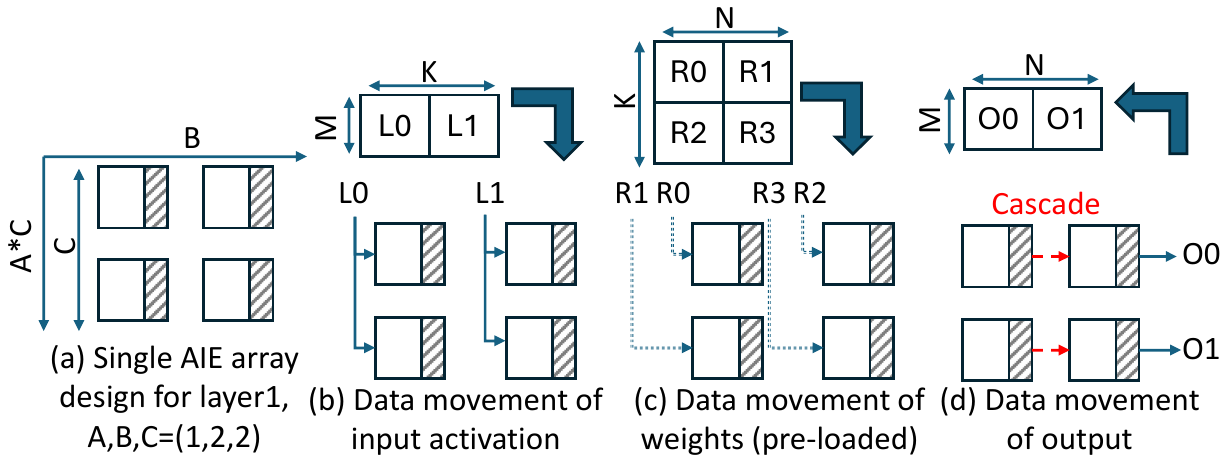}
    \vspace{-10pt}
    \caption{Data movement pattern of an example matrix multiplication layer as an AIE array.
    }
    \vspace{-10pt}
    \label{fig: single_layer_arch}
\end{figure}

On the layer level, one NN layer is implemented as one 2-D AIE array.
For an MM layer of $M\times K\times N$, we spatially partition the layer $A\times B\times C$ times along the M, K, N dimensions, respectively.
A layer can be regarded as in a rectangular layout with $A\cdot C$ rows and $B$ columns, as shown in Figure~\ref{fig: single_layer_arch}(a).
Each layer will take a partitioned piece of input (\ref{fig: single_layer_arch}(b)) and weight (\ref{fig: single_layer_arch}(c)), and generate a piece of partial results.
The partial results are then cascaded along each row (i.e., the K dimension) for (\ref{fig: single_layer_arch}(d)).
It is worth noting that after accumulation, all full results reside in the rightmost column of this AIE array in a distributed manner.  

\vspace{-10pt}
\subsection{Inter-layer Communication}
\label{sec: inter-layer communication}

\subsubsection{Pattern of Direct Communication between Layers}
\begin{figure}
    \centering
    \includegraphics[width=0.85\linewidth]{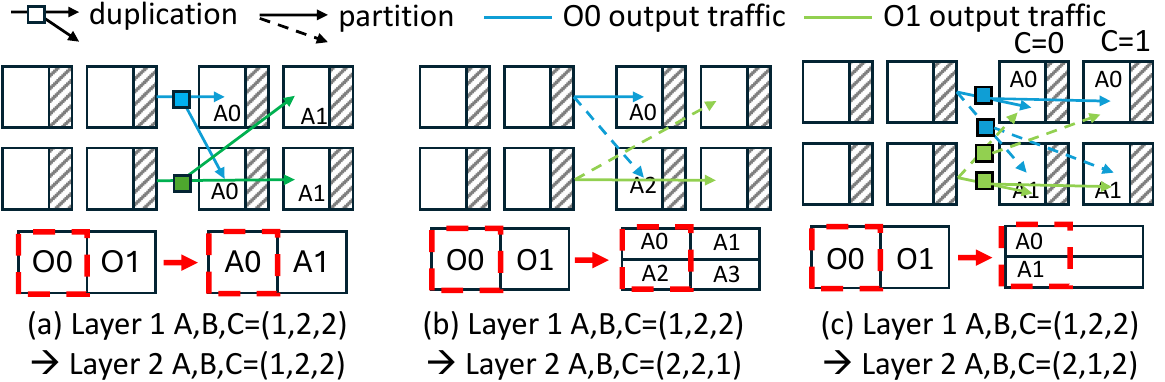}
    \vspace{-10pt}
    \caption{Inter-layer intermediate activation data communication patterns between example layers.
    }
    \label{fig: inter_layer_comm}
    \vspace{-10pt}
\end{figure}
When enabling direct communication for intermediate data, the traffic pattern between two consecutive layers changes with the spatial parallelism of these layers, as shown in Figure~\ref{fig: inter_layer_comm}.
We denote that the first layer is partitioned with $A, B, C$, and the second is with $A', B', C'$.
As shown in Figure~\ref{fig: inter_layer_comm}(a), when $C'>1$, more than one AIE tile in layer 2 takes the same input, as a result, the output piece $O_0$ will be \textbf{duplicated} to 2 AIEs which have the same $A$ and $B$, but different $C$ indexes.
In this pattern, each destination gets the same data.
In addition to duplication, when $A'>A$ or $B'>C$, the output data will need to be \textbf{partitioned}, with each piece sent to different AIEs.
An example is shown in Figure~\ref{fig: inter_layer_comm}(b), where the output $O_0$ is partitioned into two pieces along the $M$ dimension, 
In contrast, when $A'<A$ or $B'<C$, one output kernel will receive different data from different sources.
Moreover, different traffic patterns can be mixed, as shown in Figure~\ref{fig: inter_layer_comm}(c), where $A<A'$ and $C'>1$, one output will be partitioned along the M dimension, then each piece will be duplicated to two destinations.

\subsubsection{Direct DMA Connection}
\M~supports using DMA to connect two consecutive layers directly with the \texttt{IO buffer} APIs. 
The duplicated data can be passed using one IO buffer via multicasting, which is implemented by connecting the same output buffer in the source to multiple input buffers in the destination.
When data is partitioned, different IO buffers have to be used for each piece.
Subsequently, the $i,j,k$ loops in the kernel are split into multiple smaller blocks, each one reading or writing to the same buffer.
As no constraints are posed on the AIE location, the DMA connections can be conducted from the source AIE to the destination directly.

\subsubsection{Cascade Connection}
\label{sec: cascade connection}
\begin{figure}
    \centering
    \includegraphics[width=0.9\linewidth]{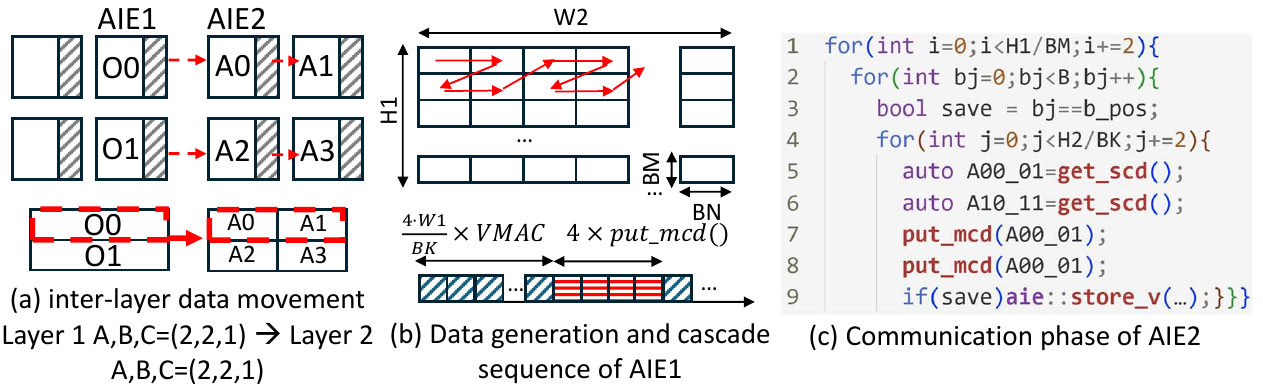}
    \vspace{-10pt}
    \caption{Cascade connection for inter-layer data movement.
    }
    \vspace{-10pt}
    \label{fig: cascade_communication}
\end{figure}
An AIE tile can cascade only to its south or east neighbor, and the AIE API limits each AIE to at most one input and one output cascade port; this poses constraints on enabling cascade connections for inter-layer intermediate data.
\M~requires the consecutive two layers to have $A=A'$  and $C=C'=1$ to enable cascade connection for inter-layer intermediate data.
Under this constraint, the inter-layer data will only need to be passed horizontally, as shown in~\ref{fig: cascade_communication}.
Moreover, the second layer should be placed directly to the east of the first layer.

The cascade is implemented as a 512-bit wide FIFO with depth = 4 in hardware.
Figure~\ref{fig: cascade_communication}(b) demonstrates the data generation on the sender side, where the AIE1 generates a piece of the output matrix of shape $H_1\times W_2$.
Each block in the figure represents an output of a \texttt{VMAC} and has the size $B_M\times B_N$.
As the \texttt{i,j} loops are unrolled by 2 in one AIE kernel, each \texttt{j} loop generates 4 blocks.
Then,~\M~uses the \texttt{put\_mcd()} to send out the data via cascade.
The blocks are pushed into the cascade sequentially, and a data layout change will be needed, for example, in INT8, one output block has 256 bits size, thus 2 blocks will be concatenated to suit the 512-bit cascade width.

On the receiver side, if $B'>1$, the intermediate data will be partitioned to different AIEs within a row.
As shown in Figure~\ref{fig: cascade_communication}(c), for each AIE within layer 2, it will receive a complete piece of output from upstream(line 1-6), and send the same complete piece downstream(line 6-7).
One AIE will only save the data corresponding to its location along the $B$ dimension within the layer(line 3,9).
After all data has been saved and transferred downstream, one AIE kernel can then load from its local memory, compute, and store.

Different from DMA or shared local memory communication, the cascade connection does not need to wait for layer 1 to finish before launching layer 2 kernels.
Instead, the computation phase of AIE1 and the communication phase of AIE2 can overlap with each other, as most inter-layer data is generated and processed when AIE1 is still computing.
This enables~\M~to reduce almost all the communication latency except for the last \texttt{j} loop.

\vspace{-10pt}
\subsection{Non-MM Kernels Design \& Implementation}

\subsubsection{Global aggregation layers:}
\M~supports global aggregation layers to support handling DeepSets models fully within the AIE array instead of sending the intermediate data to the FPGA fabric for this non-linear layer.
As a result, the communication overhead incurred can be reduced.
In a DeepSets model, the input matrix is first sent to an MLP named \textit{phi} layers and gets an $M\times F$ matrix.
The global aggregation layer is then applied, reducing the $ M$ dimension and generating a $1\times F$ vector.
Mean and sum reduction are used by the DeepSets models and supported by~\M.

\begin{figure}
    \centering
    \includegraphics[width=0.9\linewidth]{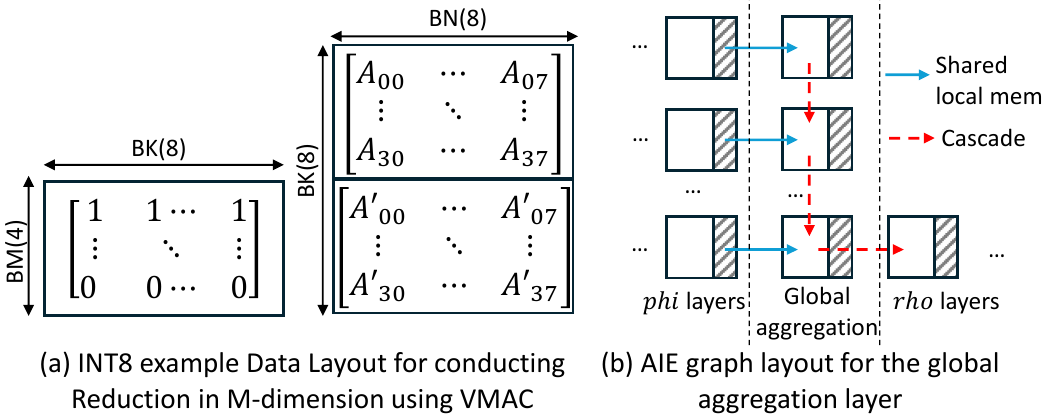}
    \vspace{-10pt}
    \caption{Global aggregation layer design.
    }
    \label{fig: global_aggregation}
    \vspace{-10pt}
\end{figure}

Two key challenges arise in designing high-performance global aggregation layers.
First, the output of the linear layers is in a blocked layout.
A block contains $B_M$ rows of the matrix and occupies one vector register.
However, AIE cannot reduce within one register directly, and moving the data to different registers and then reducing will incur a large amount of \texttt{VMOV} and \texttt{VADD} instructions in the critical path, slowing down the execution.
Moreover, as the output of the \textit{phi} layers is distributed to multiple AIEs, inter-AIE communication is inevitable, and the overhead has to be reduced.

Figure~\ref{fig: global_aggregation} demonstrate the solution of~\M. 
In the single-kernel level,~\M~conducts the reduction along the M dimension using \texttt{VMAC}.
The input of the aggregation layer is posed as the RHS, and the LHS is a fixed matrix with only the first row being all 1.
After the MAC operation, a $BM\times BN$ matrix will be generated and kept in an accumulator register, with the first row being the reduced vector.
Similarly, larger matrices of $M\times N$ can be reduced as a $B_M\times M\times N$ MM kernel.
Data layout change may be needed.
For example, in INT8, $B_K = 2\cdot B_M$, thus two output blocks $A$ and $A'$  can be concatenated and reduced in one instruction, as shown in Figure~\ref{fig: global_aggregation}(a).

In the AIE-array level, to avoid using the DMA connection, we assume that the last layer of \textit{phi} layers has $C=1$.
As shown in Figure~\ref{fig: global_aggregation}(b), in total $A$ AIEs are placed adjacent to the \textit{phi} layers for aggregation.
The shared local memory connection is used to pass the inter-layer data to the aggregation layer.
For each AIE in the aggregation layer, it first reduces the $H_1\times W_2$ matrix received from the last layer, and sends out the accumulated results after reducing each column.
The final aggregated results can be sent to the rho layers directly via cascade; if mean reduction is applied, only the last AIE needs to compute the division.

\subsubsection{Bias, ReLU, and Quantization Scheme} 
~\M~supports Bias and ReLU layers after each matrix multiplication layer.
These two layers are enabled within the last(rightmost) column of an AIE array and are integrated within the epilogue of the matrix multiplication.
The bias of each dense layer is pre-loaded to the local memory of each AIE as runtime parameters.
~\M~also supports INT8 symmetric quantization to the input activation and weights.
Integer scales are assumed to be powers of two.
Quantization happens when converting the INT32 accumulator register back to INT8 by bit-shifting.
The Bias data is stored in INT32 in the local memory.
To apply bias to a $B_M\times B_N$ block, we first load $1\times B_N$ bias data from local memory, then duplicate it to $B_M\times B_N$ by \texttt{concat} it with itself.
The duplicated bias is then added to the MM results.

\vspace{-10pt}
\section{\M~Performance Model and Design Space Exploration}
In this section, we first demonstrate the overhead-aware performance model of~\M, then discuss the design space exploration.

\subsection{Overhead-aware Performance Model}
We model the performance of~\M~as two parts: the computation latency $L_{comp}$ of each layer and the communication latency between layers $L_{comm}$.
The total latency of a model is the summation of them.

\subsubsection{Single AIE computation latency}
When compiling an AIE kernel, the \texttt{aiecompiler} tries to pipeline the instructions for load, compute, and store.
However, the prologue to launch the pipeline and the epilogue to process the output results will elongate the latency.

\M~models single AIE kernels' latency $L_{AIE}$ using:
\begin{equation}
    L_{AIE} = \frac{H_1\cdot W_2}{4\cdot B_M\cdot B_N}L_j+L_o
\end{equation}
Where $L_j$ is the latency of each \texttt{j} loop.
Within one \texttt{j} loop, the \texttt{VMAC} instructions can usually be pipelined with $II=1$ in the load-compute pipeline, then the epilogue takes some additional cycles:
\begin{equation}
    L_{j} = 4\frac{W1}{B_K}+L_{epi}
\end{equation}
The $L_o$ represents the overhead that cannot be pipelined, and is composed of the latency of load/compute/store instructions, the synchronization overhead, and the overhead incurred by imperfect pipelining in the first and last iteration.
$L_o$ varies with different implementations of communication patterns, for example, when using cascade communication, the results will not store to the local memory, and thus the latency of store instructions will not be paid.

\subsubsection{AIE array computation latency}
Different AIEs in one AIE array compute the same shape of MM, but have different epilogue operations.
Moreover, AIEs that are cascaded for partial results interfere with each other due to empty stalls or back pressure.
We model the $a$-th AIE along the $K$ dimension as follows, where $L_{cas}$ represents the latency brought cascade communication:
\begin{equation}
    L^{a}_{j} = 4\frac{W1}{B_K}+L_{epi}+L_{cas}
\end{equation}
In each \texttt{j} loop, one AIE waits for the upstream to pass the partial results, then begins computation, then sends the accumulated results downstream.
Thus, the \texttt{j} loop latency is bounded by the longest one among the cascaded AIEs, and there will be $ \frac{H_1\cdot W_2}{4\cdot B_M\cdot B_N}+B-1$ \texttt{j} loops in total:
\begin{equation}
    L_{comp} = (\frac{H_1\cdot W_2}{4\cdot B_M\cdot B_N}+B-1)\max_{a} (L^a_j)+L_o
\end{equation}

\subsubsection{Inter-layer communication latency}
The communication latency between layers depends on the method used.
When the DMA connection is applied, the communication happens sequentially with the computation.
That is, only after the producer kernel releases the output buffer (typically when the kernel finishes) can the DMA communication begin. Then the producer kernel cannot be launched until the DMA finishes.
To be noted, although double buffer is applied to DMA, it overlaps the latency of different input samples.
For the same input sample, the latency is not reduced.
The DMA latency can be characterized as three parts: the initial latency $L_{init}$ for initialization and synchronization; the data transfer latency with 32 bits/cycle throughput. In addition, it takes $4\cdot D$ cycles to pass the data from source to destination, where $D$ represents the Manhattan distance.
When two layers with multiple AIEs communicate via DMA, the longest DMA distance is used for estimation:
\begin{equation}
L_{comm}^{DMA}=L_{init}+\frac{DATA\_SIZE}{32}+4\cdot D
\end{equation}

When cascade is used, most of the communication latency is overlapped, as discussed in Section~\ref{sec: cascade connection}.
The gap between the computation phase of the producer and the consumer is a constant, which includes the producer passing the last blocks of data to the consumer, and the consumer storing it in local memory:
\begin{equation}
L_{comm}^{cas}=O_{cas}
\end{equation}

\subsection{DSE for Mapping and Placement}

\begin{figure}
    \centering
    \includegraphics[width=0.75\linewidth]{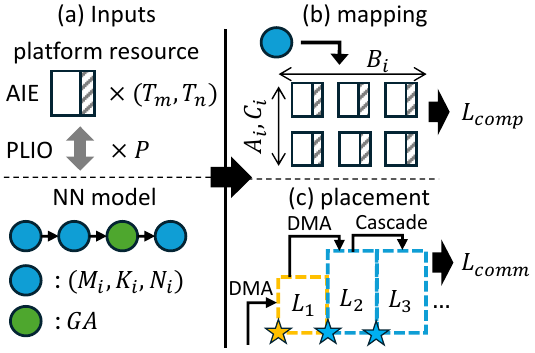}
    \vspace{-10pt}
    \caption{\M~design space exploration.    }
    \label{fig: dse}
    \vspace{-15pt}
\end{figure}

\M's design space exploration includes both mapping, i.e., deciding how many AIE kernels to use, and placement, i.e., deciding where to deploy these kernels in the AIE array.
This is because both affect the applicability of the cascade connection, which can be enabled between two layers only if (1) they are mapped with $A=A'$ and $C=C'=1$, and (2) the second layer is placed directly to the right of the first.
Figure~\ref{fig: dse} shows the workflow of~\M~design space exploration.
The inputs include platform resources, i.e., the two dimensions of the AIE array and the PLIO number, and NN model specifications, i.e., the shape of each layer.
\M~supports MLP- and DeepSets-based model architectures, where layers are assumed to form a sequence with at most one global aggregation.

~\M~first searches the mapping for all layers, as shown in Figure~\ref{fig: dse}(b).
For each layer, we brute-force search the spatial parallelism $A_i, B_i, C_i$, with respect to the total AIE tile number that $\Sigma A_i\cdot B_i\cdot C_i \leq T_m\cdot T_n$.
PLIO is used to load the data from the FPGA fabric in the first layer, and to store it in the last layer, thus we have $A_1*B_i+A_n*C_n\leq P$, where $A_n, C_n$ are from the last layer.
The MM shape for one AIE $H_1, W_1, W_2$ can be determined when $M, K, N$ and $A, B, C$ are fixed.
We require $H_1>2\cdot B_M$, $W_1> B_K$, and $W_2>2\cdot B_N$ so that there are enough data for the single AIE execution.
Further, the $A, B, C$ are constrained to powers of 2.
After this step, the computation latency can be obtained.

~\M~next places layers onto the AIE array.
As shown in Fig.~\ref{fig: dse} (c), each layer occupies a rectangular region with $(A \times C)$ rows and $B$ columns.
Layers are placed sequentially in a left-to-right, bottom-to-top order.
For each layer, the placement position is determined by selecting the bottom-left tile with the minimum row index, and among such candidates, the minimum column index.
As a result, if two consecutive layers are consistent in mapping, they can be placed adjacent, and a cascade connection can be used (e.g., $L_2$ and $L_3$).
Otherwise, the direct DMA connection is used (e.g., $L_1$ and $L_2$).
For the global aggregation layer, we constrain the linear layer before it to have $C=1$.
During mapping, the aggregation layer is always placed adjacent to that linear layer.

\vspace{-10pt}
\section{Evaluation}
\subsection{Experiment Setup}
We prototype \M~on the AMD Versal VEK280 evaluation board with the AIE-ML architecture, which contains 304 (8 rows, 38 columns) AIE tiles.
The Vitis and Vivado 2024.1 toolchain is used for compilation. 
The AIE and PL frequencies are set to 1.25 GHz and 330 MHz, respectively.
We first report the single-AIE-kernel to show~\M's implementation is comparable to the state-of-the-art works and the effectiveness of~\M's overhead-aware performance model.
We then evaluate~\M~on different synthetic and realistic MLP- and DeepSets-based workloads, and compare with the state-of-the-art AIE-based and FPGA-based frameworks, i.e., SSR, AIE4ML, and HLS4ML.
The PL frequencies for the SSR and HLS4ML baseline are set to 230MHz and 200MHz, following their default setup.
For AIE4ML, Vitis 2025.1 is used for compilation.

\vspace{-10pt}
\subsection{Single AIE Performance}
\begin{table}[t]
\centering
\caption{Single AIE computation time among different MM shapes without and with bias and ReLU, with efficiency (\%)}
\vspace{-10pt}
\footnotesize
\begin{tabular}{ccccc}
\toprule
Layer Shape & GAMA & AIE4ML(+BR) & \M & \M(+BR) \\
\midrule
16$\times$16$\times$16   & 32    & 34.4   & 31.2 (41.0\%)   & 34.4 (37.2\%) \\
32$\times$32$\times$32   & 184   & 194.4  & 129.6 (79.0\%)  & 184 (55.7\%) \\
64$\times$64$\times$64   & 897.6 & 1109.6 & 868 (94.4\%)    & 967.2 (84.7\%) \\
8$\times$32$\times$32    & 63.2  & 82.4   & 45.6 (56.1\%)   & 56 (45.7\%) \\
8$\times$64$\times$64    & 124.8 & 167.2  & 123.2 (83.1\%)  & 136 (75.3\%) \\
8$\times$128$\times$128  & 438.4 & 525.6  & 438.4 (93.4\%)  & 525.6 (77.9\%) \\
\bottomrule
\end{tabular}
\label{tab: single aie latency}
\vspace{-10pt}
\end{table}
\M's main contribution is advancing the communication between layers. 
Here, we report the single-AIE performance to show that our implementation is comparable to state-of-the-art works.
Table~\ref{tab: single aie latency} demonstrates the single AIE performance in nanoseconds and the corresponding hardware efficiency.
We choose GAMA (without Bias and ReLU) and AIE4ML (with Bias and ReLU), the state-of-the-art frameworks targeting AIE-ML architecture for comparison. 
The computation time in ns is reported; the DMA latency for load and store is omitted.
Designs with bias and ReLU are tagged with (+BR).
As shown in Table~\ref{tab: single aie latency}, ~\M~reaches comparable performance compared with state-of-the-art works.
When the workload size is small, the hardware utilization of all implementations is small, as the hardware overhead dominates the latency.
The utilization increases as the workload grows, and~\M~reaches $>$ 90\% utilization for $64\times64\times64$ and $8\times128\times128$ workloads.

\begin{figure}
    \centering\includegraphics[width=0.9\linewidth]{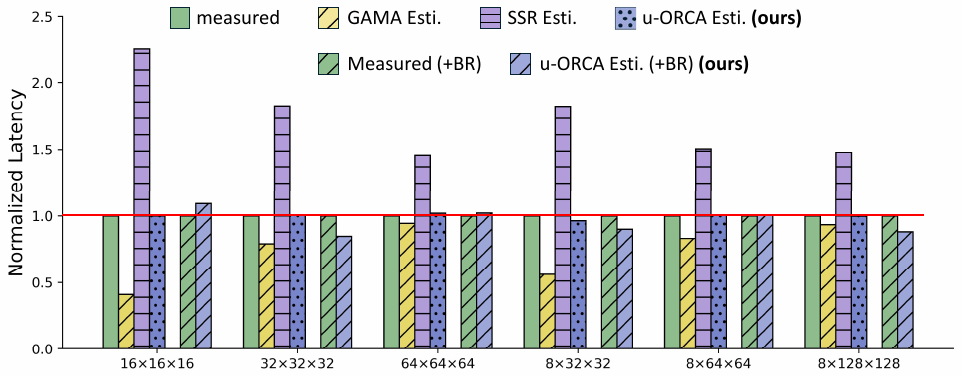}
    \vspace{-10pt}
    \caption{Normalized single AIE measured latency and estimated latency from different performance models.
    }
    \label{fig: perf model}
    \vspace{-20pt}
\end{figure}

\vspace{-10pt}
\subsection{Performance on Synthetic MLPs}
\begin{figure*}[t]
    \centering
    \includegraphics[width=\textwidth]{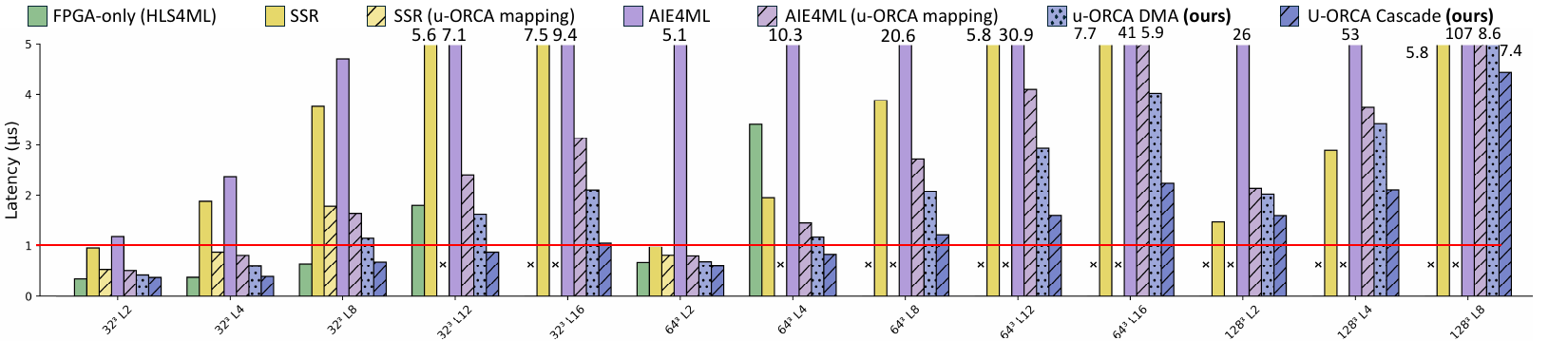}
    \vspace{-10pt}
    \caption{Latency comparison of synthetic MLP workloads with various layer shapes and number of layers.}
    \label{fig: synthetic MLP}
    \vspace{-10pt}
\end{figure*}

We evaluate~\M~on synthetic MLP workloads composed of square matrix multiplications with varying layer counts and matrix sizes.
We use three representative baselines: HLS4ML (PL compute and PL inter-layer communication), SSR (AIE compute and PL inter-layer communication), and AIE4ML (AIE compute and shared memory tile inter-layer communication).
To isolate the effectiveness of the proposed inter-layer cascade communication, we also implement SSR and AIE4ML variants using the same mapping and placement as~\M~Cascade.
As an ablation, we include~\M-DMA, which uses AIE for computation and passes intermediate data directly between layers without entering the memory tile.
We profile end-to-end latency, measured from the first input element entering the PLIO to the last element leaving the PLIO.

Figure~\ref{fig: synthetic MLP} compares latency across frameworks.
Each cluster $s^3 Ly$ reports the latency of an MLP with $y$ layers, where each layer is an $s\times s\times s$ MM kernel.
\M~Cascade significantly outperforms the baselines, achieving average latency reductions of 1.7$\times$, 3.9$\times$, 7.6$\times$, and 1.4$\times$ over feasible HLS4ML, SSR, AIE4ML, and~\M-DMA designs, respectively.
It further achieves $1.91\times$ and $1.95\times$ reductions over SSR and AIE4ML using~\M~mapping.

For HLS4ML, the reuse factor balances latency and utilization.
For tiny workloads, such as two to four layers of size 32, HLS4ML keeps all computation on chip and achieves slightly lower latency than~\M.
However, as the input size grows, resource usage quickly saturates; high reuse factors then cause much higher latency (e.g., $64^3L4$), or compilation fails because utilization remains $>1$ even with a reuse factor of 32 (e.g., $64^3L8$ and above).
For SSR, with few layers, variants using our mapping reduce latency by assigning more AIE kernels to each layer.
As the number of layers increases, they fail to compile due to insufficient PLIO ports, since~\M~spatially maps all layers to different accelerators.
In contrast, the original SSR maps multiple layers to the same accelerator and thus supports more layers.
For $128^3L2$, SSR outperforms~\M~because it uses 64 AIE kernels as a $4\times4\times4$ array, whereas~\M~is limited by the $C=1$ cascade constraint and uses only 32 AIE kernels as an $8\times4\times1$ array.
With more layers,~\M~can use more AIEs and again outperforms SSR.
For AIE4ML, the default design assigns one AIE per layer, leading to long latency.
Applying~\M~mapping assigns more AIEs and improves performance, but latency remains higher than~\M~due to inefficient inter-layer communication.

When considering a 1-$\mu$s latency budget, ~\M~supports $>12$ layers of size 32 MMs, or $>4$ layers of size 64 MMs.
One major component of latency for~\M~is the DMA communication before the first layer and after the last layer.
For the $64^3L8$ MLP, it reaches 1.21 $\mu$s, whereas the DMA takes 0.3 $\mu$s, taking roughly 25\% of the end-to-end latency.
On the other hand, the baselines can support only up to 8 layers of size 32, or 2 layers of size 64 MMs.
\vspace{-10pt}
\subsection{Performance on Realistic MLPs}
\begin{table}[t]
\centering
\footnotesize
\caption{Realistic MLP and Deepsets workloads used.}
\vspace{-12pt}
\begin{tabular}{lcc}
\toprule
Model & Input & \# Nodes per Layer \\
\midrule
JSC-M & $64 \times 16$ & [64, 32, 32, 32, 5] \\
JSC-XL & $64 \times 16$ & [128, 64, 64, 64, 5] \\
JSC-XL-d & $64 \times 16$ & [128, 128, 64, 64, 64, 64, 64, 5] \\
\midrule
Deepsets-32 & $32 \times 21$ & phi:[32, 32, 32], rho:[32, 10] \\
Deepsets-64 & $64 \times 21$ & phi:[64, 64, 64], rho:[64, 10] \\
Deepsets-32-d & $32 \times 21$ & phi:[32, 32, 32, 32, 32], rho:[32, 10] \\
Deepsets-64-d & $64 \times 21$ & phi:[64, 64, 64, 64, 64], rho:[64, 10] \\
\bottomrule
\end{tabular}
\label{tab: realistic workloads}
\vspace{-13pt}
\end{table}

\begin{figure}
\centering\includegraphics[width=0.9\linewidth]{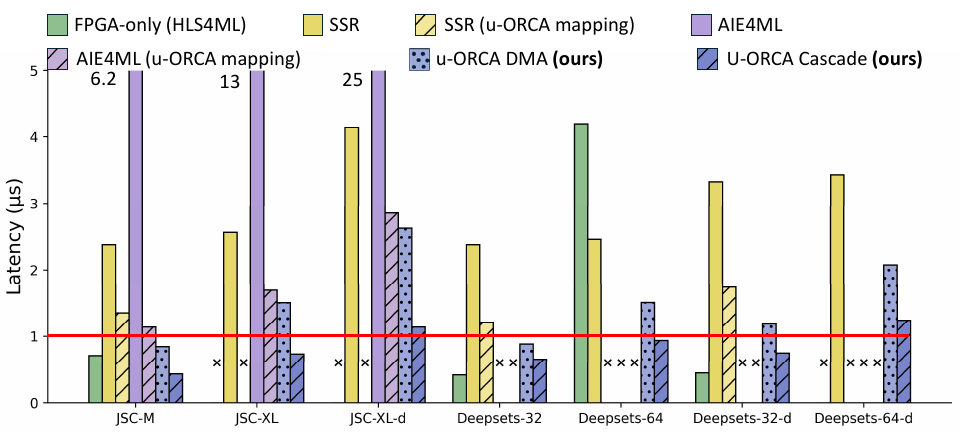}
    \vspace{-15pt}
    \caption{Latency comparison on realistic workloads
    }
    \label{fig: real MLP exp original mapping}
    \vspace{-12pt}
\end{figure}

We also evaluate~\M~on realistic MLP- and DeepSets-based models.
The model details are as shown in Table~\ref{tab: realistic workloads}. For the DeepSets model, the input is first processed by the phi layers, and the one global aggregation layer computes the output of the phi layers along the M dimension. The $1\times32$ or $1\times64$ ouput are then sent to the rho layers. 
As AIE-ML does not support global aggregation, the corresponding DeepSets performance is omitted.
~\M~reaches 1.83$\times$, 3.75$\times$, 18.33$\times$, and 2.09$\times$ latency reduction over the HLS4ML, SSR, AIE4ML, and~\M~DMA designs, and 2.42$\times$, 2.47$\times$ over SSR and AIE4ML with~\M mapping.
The higher gain on realistic over synthetic MLP workloads is because (1) the realistic MLPs usually have more layers, (2) the realistic MLPs reduce \#nodes for the last layer, which decreases the output DMA latency.
Among all 7 realistic workloads,~\M~reaches the 1 $\mu$s latency for 6 of them, with the DeepSets-64-d taking 1.1 $\mu$s.
HLS4ML can achieve this latency budget in the smallest 3 workloads, ~\M~DMA achieves in 2 of them, and SSR and AIE4ML cannot reach this budget.

\vspace{-10pt}
\subsection{Performance on Global Aggregation layers}
\begin{table}[t]
\centering
\footnotesize
\caption{Latency of different global aggregation layers.}
\vspace{-12pt}
\begin{tabular}{ccccc}
\toprule
Input & \#AIE & Baseline & Ours & Speedup \\
\midrule
$32 \times 32$ & 4 & 373 & 66 & 5.65$\times$ \\
$32 \times 64$ & 4 & 760 & 72 & 10.56$\times$ \\
$64 \times 32$ & 8 & 397 & 139 & 2.86$\times$ \\
$64 \times 64$ & 8 & 834 & 145 & 5.75$\times$ \\
\bottomrule
\end{tabular}
\label{tab: global aggregation exp}
\vspace{-17pt}
\end{table}
\vspace{-4pt}
The latency of the global aggregation layer with different shapes and AIE numbers is shown in~\ref{tab: global aggregation exp}.
Since existing frameworks do not implement the global aggregation layer, we provide an in-house baseline.
In this baseline, the inter-kernel connection is the same as our MAC-based implementation; within each AIE kernel, the reduction is computed by using the \texttt{extract()} API to get each row in the output block, using \texttt{aie::add} to reduce different rows, and using \texttt{insert()} to put the result into an accumulator register, which is then sent to the downstream AIEs.
This baseline incurs many vector movement operations in the compiled instruction sequence, which cannot be overlapped.
As a result, single-kernel operations dominate the latency: the latency increases largely with larger matrix per AIE and slightly with more AIEs used.
Compared with the baseline,~\M~MAC-based approach uses only one MAC operation for computation, achieving at least 2.8$\times$ speedup.
Its main latency comes from memory and cascade overhead, so the latency increases with more AIEs used.
\vspace{-10pt}
\subsection{~\M~Performance model effectiveness}
\vspace{-4pt}
Figure~\ref{fig: perf model} compares normalized measured and estimated latency for different single-AIE workloads.
For each shape, the two subclusters report latency without and with bias and ReLU, respectively.
We use GAMA and SSR as baselines.
GAMA relies on theoretical cycle counts and is therefore over-optimistic, while SSR uses profile-based parameters that work well for larger workloads and within an AIE array but are over-pessimistic for small workloads.
For single-AIE workloads,~\M~achieves 1.1\% average error without bias and ReLU and 4.6\% across all workloads, compared with 25.5\% and 72.3\% for GAMA and SSR, respectively.
For end-to-end models,~\M~achieves 15.6\% average error.

\smallskip
{\small
{\noindent\textbf{ACKNOWLEDGEMENTS --}} This work was partially supported by Brown University New Faculty Start-up Grant, DOE award DE-SC0026344,
NSF awards 
\#2348306, 
\#2511445, 
\#2518375, 
\#2536952, 
\#2544032. 
We thank AMD for the hardware and software donations.
P. Zhou has a financial interest in Shanghai Suikun.

\bibliographystyle{ACM-Reference-Format}
\balance
\bibliography{ref_short}

\end{document}